\documentclass[12pt,twocolumn,preprint]{aastex6}

\usepackage{CJK}
\usepackage{color}

\usepackage{natbib}
\usepackage{hyperref}
\usepackage{cleveref}
\usepackage{url}

\begin{document}
\title{ GW190814's secondary component with mass $(2.50-2.67)$ M$_{\odot}$ as  a super-fast pulsar}
\author{Nai-Bo Zhang\altaffilmark{1} and Bao-An Li\altaffilmark{2}$^{*}$}
\altaffiltext{1}{Shandong Key Laboratory of Optical Astronomy and Solar-Terrestrial Environment, School of Space Science and Physics, Institute of Space Sciences, Shandong University,  Weihai, Shandong, 264209, China\\}
\altaffiltext{2}{Department of Physics and Astronomy, Texas A$\&$M University-Commerce, Commerce, TX 75429, USA\\
\noindent{$^{*}$Corresponding author: Bao-An.Li@Tamuc.edu}}

\begin{abstract}
Using Stergioulas's RNS code for investigating fast pulsars with Equation of States (EOSs) on the causality surface (where the speed of sound equals that of light) of the high-density EOS parameter space satisfying all known constraints from both nuclear physics and astrophysics, we show that one possible explanation for the GW190814's secondary component of mass $(2.50-2.67)$ M$_{\odot}$ is that it is a super-fast pulsar spinning faster than 971 Hz about 42\% below its Kepler frequency. If confirmed, it would be the fastest pulsar with the highest mass observed presently. There is a large and physically allowed EOS parameter space below the causality surface where pulsars heavier than 2.50 M$_{\odot}$ are supported if they can rotate even faster with critical frequencies depending strongly on the high-density behavior of nuclear symmetry energy.
\end{abstract}
\maketitle

\section{Introduction}
The recent  LIGO/Virgo observation of GW190814 from the merger of a black hole (BH) of mass (22.2-24.3) M$_{\odot}$ and a secondary compact object $m_2$ with mass $(2.50-2.67)$ M$_{\odot}$ provided an exciting new stimulus to the ongoing debate on whether/where a gap exists between the maximum mass of neutron stars (NSs) and the minimum mass of BHs \citep{LV20}. The highly unequal masses of the two objects involved and the unusually small secondary mass make the source of GW190814 unlike any other compact binary coalescence observed so far. As discussed in detail in the LIGO/Virgo discovery paper \citep{LV20}, the nature of GW190814's secondary component is largely unknown as no evidence of measurable tidal effects in the signal and no electromagnetic
counterpart to the gravitational waves were identified. It is thus not clear if the $m_2$ is a BH, NS, or other exotic objects.

Already several interesting proposals have been made \citep[see, e.g.,][]{LV20,Most,Fishbach,Tan,Leh,Vattis,Essick,Zevin,Harvard,Armen}. Since it is well known that rotations provide additional support to the pressure balancing the gravity, leading to a NS maximum mass at the Kepler frequency about 20\% higher than that of the static NS for a given nuclear Equation of State (EOS)
\citep[see, e.g.,][]{Cook1994,Lasota1996,Lattimer-sci,Plamen08,Haensel2008,Haensel2009,Breu,Wei17}, the possibility for the GW190814's secondary as a rapidly rotating NS was first studied by the LIGO/Virgo Collaboration \citep{LV20}. Since the spin parameter of the secondary was not observationally constrained and the calculation of the NS maximum mass depends on the unknown EOS of super-dense neutron-rich nuclear matter, conclusions regarding rotational effects on GW190814's secondary mass are not clear. In \citet{LV20}, taking $2.3$ M$_{\odot}$ as the maximum mass M$_{\rm TOV}$ of non-rotating NSs based on estimates from studying the merger remnant of GW170817,  it was found that {\it although the degree of EOS uncertainty is difficult to quantify precisely if we take the more conservative $2.3$ M$_{\odot}$ bound at face value, then $m_2$ is almost certainly not an NS}. On the contrary,  \citet{Most} also adopted M$_{\rm TOV}=2.3$ M$_{\odot}$ in a more detailed study using universal relations connecting the masses and spins of uniformly rotating neutron stars \citep{Breu}, it was found that {\it the secondary $m_2$ does not need to be an {\it ab-initio} BH nor an exotic object; rather, it can be a rapidly rotating neutron star that collapsed to a rotating BH at some point before the merger. Moreover, a new bound of M$_{\rm TOV}\ge 2.08\pm 0.04$ M$_{\odot}$ was obtained even in the less likely scenario in which the secondary NS never collapsed to a BH. }

While it is probably more interesting to study all other more exotic possibilities, the existing controversy calls for further studies about the GW190814's secondary simply as a rapidly rotating NS and how fast it really has to rotate with respect to its Kepler frequency. In this work, using Stergioulas's RNS code for investigating rapidly rotating compact stars \citep{RNS}, we study the minimum frequency $f_{2.5}$ that can rotationally support an NS of mass 2.50 M$_\odot$ (and the corresponding spin parameter $\chi_{2.5}$)
within the high-density EOS parameter space bounded by the NS tidal deformability from GW170817 and radii of canonical NSs from X-ray observations using Chandra, XMM-Newton and NICER as well as nuclear theories and experiments. On the causality surface where the EOSs are the stiffest physically possible, the minimum value of $f_{2.5}$ is 971 Hz while the ratio of $f_{2.5}$ over Kepler frequency $f_K$, i.e.,  $f_{2.5}/f_K$, is between 0.578 and 0.876 (the corresponding $\chi_{2.5}$ between 0.375 and 0.550) depending on the high-density behavior of nuclear symmetry energy. Below the causality surface,  there is a large and physically allowed EOS parameter space where the secondary of GW190814 can sustain masses above $2.50$ M$_\odot$ if they rotate even faster than those on the causality surface. Thus, within the existing bounds on the EOS from both astrophysics and nuclear physics, the GW190814's secondary component can be a super-fast pulsar spinning faster than the currently known fastest pulsar PSR J1748-2446ad of frequency 716 Hz \citep{Hessels2006}, supporting the findings of \cite{Most}.

\section{An explicitly isospin dependent EOS-generator for neutron stars at $\beta$ equilibrium}\label{EOSmodel}
Here we summarize the main features of an EOS-generator for NSs consisting of neutrons, protons, electrons, and muons (the $npe\mu$ model).  More details of our approach and its applications can be found in
refs. \citep{Zhang18,Zhang19,Zhang19a,Zhang19b,Zhang2020,Xie19,Xie20}.  Unlike the widely used spectral EoS and other similar piecewise parameterizations that directly parameterize the pressure as a function of energy or baryon density,  we start from parameterizing the energy per nucleon in symmetric nuclear matter (SNM) $E_{0}(\rho)$ and nuclear symmetry energy $E_{\rm{sym}}(\rho)$ according to
\begin{eqnarray}\label{E0para}
  E_{0}(\rho)&=&E_0(\rho_0)+\frac{K_0}{2}(\frac{\rho-\rho_0}{3\rho_0})^2+\frac{J_0}{6}(\frac{\rho-\rho_0}{3\rho_0})^3,\\
  E_{\rm{sym}}(\rho)&=&E_{\rm{sym}}(\rho_0)+L(\frac{\rho-\rho_0}{3\rho_0})\nonumber\\
  &+&\frac{K_{\rm{sym}}}{2}(\frac{\rho-\rho_0}{3\rho_0})^2
  +\frac{J_{\rm{sym}}}{6}(\frac{\rho-\rho_0}{3\rho_0})^3\label{Esympara}
\end{eqnarray}
where $E_0(\rho_0)=-15.9 \pm 0.4$ MeV \citep{Brown14} is the binding energy and $K_0\approx 240 \pm 20$ MeV \citep{Shlomo06,Piekarewicz10,Garg18} is the incompressibility at the saturation density $\rho_0$ of SNM, while $E_{\rm sym}(\rho_0)=31.7\pm 3.2$ MeV is the magnitude and $L\approx 58.7\pm 28.1 $ MeV is the slope of symmetry energy at $\rho_0$ \citep{Li13,Oertel17}, respectively. The $K_{\rm{sym}}$, $J_{\rm{sym}}$, and $J_0$ are parameters characterizing the EOS of super-dense neutron-rich nuclear matter. In particular, the $J_0$  and $J_{\rm{sym}}$ reflect respectively the stiffness of SNM EOS and nuclear symmetry energy at densities above twice the saturation density of nuclear matter. They are parameters to be inferred from astrophysical observables and/or terrestrial experiments either using the direct inversion technique \citep{Zhang18,Zhang19,Zhang19a,Zhang19b} or the Bayesian statistical approach \citep{Xie19,Xie20}. The $E_{0}(\rho)$ and $E_{\rm{sym}}(\rho)$ are then used to first construct the average nucleon energy $E(\rho ,\delta )$
in nuclear matter at nucleon density $\rho=\rho_n+\rho_p$ and isospin asymmetry $\delta\equiv (\rho_n-\rho_p)/\rho$ according to the isospin-parabolic approximation for the EOS of neutron-rich nuclear matter \citep{Bom91}
\begin{equation}\label{eos}
E(\rho ,\delta )=E_0(\rho)+E_{\rm{sym}}(\rho )\cdot \delta ^{2} +\mathcal{O}(\delta^4).
\end{equation}
The pressure in the $npe\mu$ matter core of NSs is then calculated from
\begin{equation}\label{pressure}
  P(\rho, \delta)=\rho^2\frac{d\epsilon(\rho,\delta)/\rho}{d\rho},
\end{equation}
where $\epsilon(\rho, \delta)=\epsilon_n(\rho, \delta)+\epsilon_l(\rho, \delta)$ denotes the energy density. The $\epsilon_n(\rho, \delta)$ and $\epsilon_l(\rho, \delta)$ are the energy densities of nucleons and leptons, respectively.  The core EOS is connected to the NV EOS \citep{Negele73} for the inner crust and the BPS EoS \citep{Baym71b} for the outer crust. The crust-core transition density and pressure are determined consistently from the same parametric EOS for the core. In particular, the density dependence of nuclear symmetry energy pays a very important role in determining the crust-core transition \citep[see, e.g.,][for a recent review]{BALI19}.

As discussed in detail in \citet{Zhang18,Zhang19,Xie20}, the parameterizations of both the SNM EOS $E_{0}(\rho)$ and nuclear symmetry energy $E_{\rm{sym}}(\rho)$ were chosen purposely as if they are Taylor expansions of some known energy density functions, while they are really just parameterizations and the parameters are not derivatives of some known functions but to be inferred from data. Since the parameterizations become Taylor expansions of some functions asymptotically as the density approaches $\rho_0$, this choice allows us to use nuclear theory predictions and terrestrial nuclear experiments for the EOS parameters around $\rho_0$ as guidances in setting the prior ranges and probability distribution functions (PDFs) in inferring their posterior PDFs from the observed data. Compared to directly parameterizing the normally composition-blind pressure in NSs as a function of energy or baryon density, the EOS-generator described above has the advantage of tracking explicitly the composition of the $npe\mu$ matter in NSs. For instance, with the information about the symmetry energy, one can find easily the density profile of isospin asymmetry $\delta(\rho)$ (or the corresponding proton fraction $x_p(\rho)$) at density $\rho$ through the $\beta$ equilibrium condition
$
  \mu_n-\mu_p=\mu_e=\mu_\mu
$
and the charge neutrality condition
$  \rho_p=\rho_e+\rho_\mu
$
for the proton density $\rho_p$, electron density $\rho_e$, and muon density $\rho_{\mu}$, respectively.
While the chemical potential of particle $i$ can be calculated from
$
  \mu_i=\frac{\partial\epsilon(\rho,\delta)}{\partial\rho_i}.
$

\begin{figure}
  \centering
  \includegraphics[width=8cm]{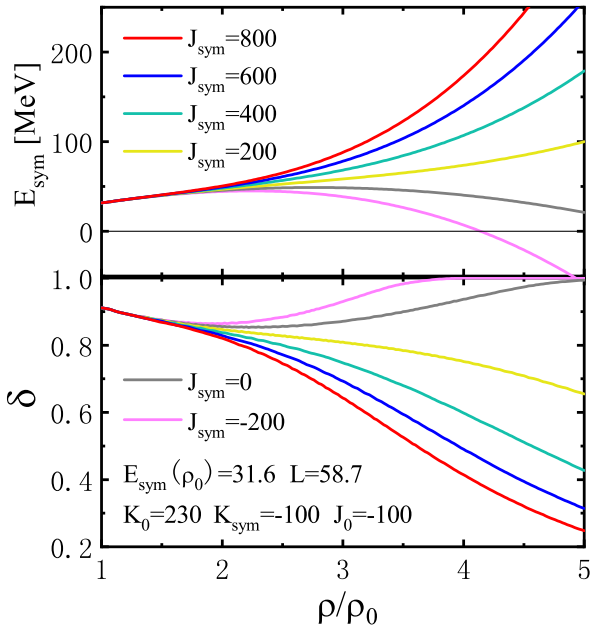}\\
  \caption{The high-density behavior of nuclear symmetry energy $E_{\rm{sym}}(\rho)$ (upper window) and the corresponding isospin asymmetry profile $\delta(\rho)$ in NSs at $\beta$ equilibrium (lower window) as functions of the reduced baryon density $\rho/\rho_0$ by varying the $J_{\rm{sym}}$ parameter within its broad range predicted by nuclear theories while all other EOS parameters are fixed at the values indicated. Taken from \citet{Zhang18}.}\label{Esym}
\end{figure}

As an example, shown in Figure \ \ref{Esym} are the high-density symmetry energy $E_{\rm{sym}}(\rho)$ (upper window) and the corresponding isospin asymmetry profile $\delta(\rho)$ in NSs at $\beta$ equilibrium (lower window) as functions of the reduced baryon density $\rho/\rho_0$ by varying the $J_{\rm{sym}}$ parameter within its broad range predicted by nuclear theories while all other EOS parameters are fixed. It is seen that the effects of varying the $J_{\rm{sym}}$ only become important above about twice the saturation density. As the $J_{\rm{sym}}$ changes from $-200$ MeV to $+800$ MeV, the symmetry energy changes from being super-soft to super-stiff. The corresponding isospin profile $\delta(\rho)$ goes from very neutron-rich or $\delta=1$ (pure neutron matter) with the super-soft $E_{\rm{sym}}(\rho)$ to almost zero (symmetric nuclear matter) with the super-stiff $E_{\rm{sym}}(\rho)$ at super-high densities. This is well understood from minimizing the $E_{\rm{sym}}(\rho )\cdot \delta ^{2}$ term in the average nucleon energy of Eq. (\ref{eos}). For easy of our following discussions, it is useful to emphasize that the symmetry energy term may contribute significantly to the total pressure. It is known that the total pressure around $2\rho_0$ has strong or even dominating contributions from the symmetry energy \citep{Lattimer00,LiSteiner,LCK08}, making the radii of canonical NSs depend strongly on the $E_{\rm{sym}}(\rho)$ around $2\rho_0$.  At even higher densities, when the $E_{\rm{sym}}(\rho)$ is super-soft, the $\delta$ is close to 1 as shown in Figure \ref{Esym}, making the $J_{\rm{sym}}$ term contribution to the total pressure as strong as the $J_0$ term in the SNM EOS $E_{0}(\rho)$ \citep{Xie19}. Consequently, the mass-radius curve and the maximum mass of NSs are very sensitive to the high-density behavior of nuclear symmetry energy \citep{BALI19}. This also explains the findings that the radii and/or tidal deformability of canonical NSs only constrain the $L$ and $K_{\rm sym}$ parameters characterizing the $E_{\rm{sym}}(\rho)$ around $(1-2)\rho_0$ but not the $J_{\rm{sym}}$ parameter \citep{Zhang18,Xie19}. To constrain the latter, one has to study the mass-radius correlations of NSs as massive as possible \citep{Xie20}. Moreover, even for rapidly rotating NSs, it has been shown earlier using the RNS code that the mass-radius sequence, the moment of inertia, and ellipticity all strongly depend on the high-density behavior of nuclear symmetry energy \citep{Plamen08,Aaron,Plamen08b}. It is thus more useful to construct the EOS of NS matter by explicitly considering the isospin asymmetry at the nucleon energy level instead of directly parameterizing the pressure as a function of energy/baryon density.

The explicitly isospin-dependent NS EOS-generator described above has been used successfully in solving the NS inverse-structure problems in both the direct inversions of NS observables in the three-dimensional (3D) high-density EOS parameter space \citep{Zhang18,Zhang19,Zhang19a,Zhang19b,Zhang2020} or Bayesian statistical inferences of multiple EOS parameters from observational data \citep{Xie19,Xie20}. It is very efficient in generating multi-million EOSs in the allowed EOS parameter space as inputs for solving the Tolman-Oppenheimer-Volkov (TOV) NS structure equations \citep{Tolman34,Oppenheimer39} in the inversion processes.

The NS EOS-generator is a numerical realization of the EOS meta-model defined by the equations given above. The parameters are randomly generated in the inversion processes in large ranges covering most if not all known predictions based on extensive surveys of nuclear many-body theories \citep{Tews17,Zhang17}. This class of EOSs is thus very general. It is also most conservative and has broad applications especially for the purpose of constraining the high-density behavior of nuclear symmetry energy described by the parameter $J_{\rm{sym}}$. The latter has been broadly recognized as among the most important but undetermined quantities affecting properties of dense neutron-rich nuclear matter \citep{Lattimer-sci,ditoro,Steiner05,LCK08,Tsang12,Chuck14,Bal16,Li17,Herman17}. The origin of the uncertain high-density symmetry energy can be traced back to our poor knowledge about the spin-isospin dependence of the many-body (three or more nucleon interactions) nuclear forces and the isospin dependence of short-range nucleon-nucleon correlations induced by the tensor force or repulsive core in dense neutron-rich matter \citep[see, e.g.,][for a recent review]{Li18}. 

In fact, to determine the high-density behavior of nuclear symmetry energy was identified as a major scientific thrust for nuclear astrophysics in both the U.S. 2015 Long Range Plan for Nuclear Sciences \citep{LRP2015} and the Nuclear Physics European Collaboration Committee (NuPECC) 2017 Long Range Plan \citep{NuPECC}. In particular, several dedicated experiments, \citep[see, e.g.,][]{russ11,Hong14,Tamii14,Xiao14,UFRIB,Wolfgang,Tsang20}, have been planned to pin down the high-density nuclear symmetry energy at the Facility for Antiproton and Ion Research (FAIR) in Europe, Facility of Rare Isotope Beams (FRIB) in the USA, Rare Isotope Beam (RIB) facility at RIKEN in Japan, High-Intensity Heavy Ion Accelerator Facility (HIAF) in China and the Rare Isotope Science Project (RISP) in Korea. The EOS of Eq. (\ref{eos}) is a basic input for transport model simulations of these experiments. It is thus critically important for the nuclear physics community. 

Determining the EOS of super-dense neutron-rich matter is a longstanding and shared goal of both astrophysics and nuclear physics \citep{Danielewicz02,Lattimer-sci,Ozel16,Watts16,Oertel17,BALI19}. As we shall illustrate with an example in the next section, the constraints on the symmetry energy around $(1-2)\rho_0$ from analyzing properties of canonical NSs using the EOS outlined above can already help rule out many predictions based on several other classes of EOS models. We thus expect the class of EOSs studied in this work is of significantly general interest to both astrophysics and nuclear physics communities.

The EOS-generator described above also has its limitations and drawbacks. Assuming the cores of NSs are made of only $npe\mu$ matter even in the possibly most massive NSs, it lacks the physics associated with the possible
phase transitions to exotic states of matter and/or productions of new particles, such as hyperons, mesons, and $\Delta(1232)$ resonances,  proposed in the literature. The appearance of new phases and particles is known to generally soften the EOS. Nevertheless, the EOS of $npe\mu$ matter serves as a useful baseline for future studies incorporating the possible new phases and particles. The necessary rotational frequency calculated within the $npe\mu$ model can be generally considered as the minimum value as a softer EOS will require a higher frequency to support a given pulsar.

We note here that the class of NS EOSs outlined above is extensible to include both new particles and the quark phase at high densities. In fact, a recent work connecting this class of NS EOS with a quark phase described with a constant speed of sound EOS within a Bayesian framework will be reported elsewhere \citep{XL20}. However,  it is well known that the critical densities for forming hyperons \citep{Toki,Lee,Kubis1,Pro19},  $\Delta(1232)$ resonances \citep{Italy,Cai-D,Ang,India,Ramos}, kaon condensation \citep{Lee,Kubis04} and the quark phase \citep{Ditoro2,Wu19} all depend sensitively on the high-density nuclear symmetry energy. For example, it has been shown extensively that because the $\Delta(1232)-\rho$ meson isovector coupling is completely unknown, not only the critical densities for forming $\Delta(1232)$ resonances in NSs but also their influences on the NS EOS are extremely uncertain \citep{Italy,Cai-D,Ang,India,Ramos}. Thus, incorporating any new particle and/or phase inevitably introduces more uncertain model parameters. 

Since the main goal of this work is to examine if the secondary component of GW190814 can be explained as a super-fast pulsar in the minimum model of NSs with the least number of uncertain model parameters, the NS EOS generator described above serves this purpose well. As we discussed in the introduction, such studies using mostly known physics as much as possible and model parameters largely under control are useful for determining whether the GW190814's secondary component is the most massive NS or the lightest BH discovered so far before resorting to less known, exotic and completely new physics. For the same reason, there is no need to introduce a fourth-order term in parameterizing the symmetry energy before the $J_{\rm{sym}}$ parameter is better determined. As we discussed above, there is currently no meaningful constraint on $J_{\rm{sym}}$ from neither astrophysical observations nor terrestrial experiments. Nevertheless, there are indeed interesting proposals to use the radii of massive NSs \citep{Xie20} or nuclear reactions induced by high energy radioactive beams \citep{FRIB} to better constrain the $J_{\rm{sym}}$ parameter. Therefore, if confirmed as an NS, GW190814's secondary component may help test the astrophysical proposals. We will thus examine some properties of the GW190814's secondary component as functions of the $J_{\rm{sym}}$ parameter in Section \ref{prop} without worrying about the possible higher-order terms.

\section{The constrained high-density EOS parameter space for massive neutron stars}
Here we illustrate the high-density EOS parameter space $K_{\rm{sym}}$-$J_{\rm{sym}}$ -$J_0$ constrained by existing astrophysical observables and the causality condition.
Much efforts have been devoted in recent years to constraining the high-density EOS parameters $K_{\rm{sym}}$, $J_{\rm{sym}}$, and $J_0$ using both terrestrial experiments and astrophysical observations \citep[see, e.g.,][for a comprehensive review]{Tesym}. Unfortunately, they are still not well determined. As we shall illustrate, the high-density SNM EOS parameter $J_0$ has the strongest control over the maximum mass of NSs. While the
high-density symmetry energy parameters $K_{\rm{sym}}$ and $J_{\rm{sym}}$ mostly control the radii, tidal deformabilities, and proton fractions of massive NSs, they also have some significant influences on the maximum mass of NSs. On the other hand, while the $L$ and $K_{\rm{sym}}$ both play significant roles in determining the radii of especially canonical NSs, they have little effects on the maximum mass of NSs. These features have been well demonstrated by many calculations using various nuclear theories and used in extracting them from astrophysical observations. However, due to the limited data available, large uncertainties still exist especially for the three high-density EOS parameters $K_{\rm{sym}}$, $J_{\rm{sym}}$, and $J_0$. For instance, using the combined data of NS tidal deformability from GW170817 and the simultaneous measurement of mass and radius of PSR J0030+0451 by the NICER Collaboration, a very recent Bayesian analysis inferred the most probable value of $K_{\mathrm{sym}}$ as $-120_{-100}^{+80}$ MeV at 68\% confidence level \citep{Xie20}. Obviously, its uncertainty is still very large. Since  the available data from canonical NSs with masses around $1.4$ M$_{\odot}$ reflect mostly the EOS around $2\rho_0$ while the $J_{\rm{sym}}$ characterizes the symmetry energy at higher densities, they do not provide much constrain on the  $J_{\rm{sym}}$ \citep{Xie19,Xie20}.  As a result, the symmetry energy at twice the saturation density is only loosely constrained to $E_{\mathrm{sym}}(2\rho_0)=54.8_{-19}^{+8.4}$ MeV at 68\% confidence level, while its behavior at higher densities is currently completely unconstrained as shown in Figure \ref{Esym}. This is well understood as we explained earlier. In the following studies, we will just use the full range of $-200 \leq J_{\rm{sym}}\leq 800$ MeV predicted by many kinds of nuclear many-body theories \citep[see, e.g.,][]{Tews17,Zhang17} for surveys of model predictions for $J_{\rm{sym}}$.

\begin{figure}
  \centering
  \includegraphics[bb=14 65 563 591, width=8cm]{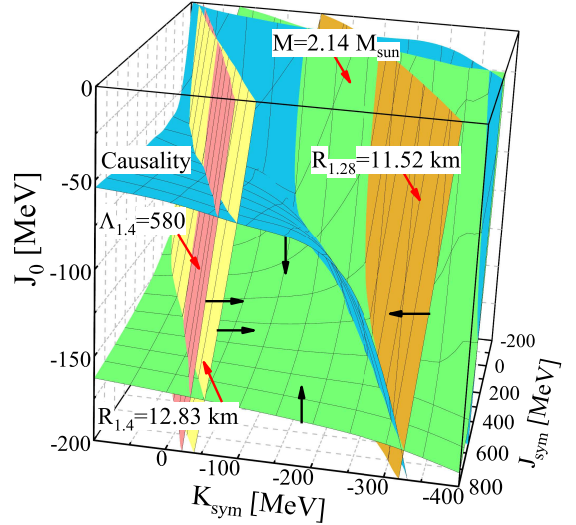}\\
  \caption{The constraints of M = $2.14$ M$_\odot$  (green surface), $R_{1.4} = 12.83$ km (yellow surface), $\Lambda_{1.4} = 580$ (red surface), $R_{1.28} = 11.52$ km (orange surface),and causality condition (blue surface) in the 3D parameter space of $K_{\rm sym}-J_{\rm sym}-J_0$. The black arrows show the directions supporting the corresponding constraints and the red arrows direct to the corresponding surfaces. Modified from Figure 4 of \citet{Zhang2020}.}\label{Constraints}
\end{figure}

\begin{figure}
  \centering
  \includegraphics[width=8cm]{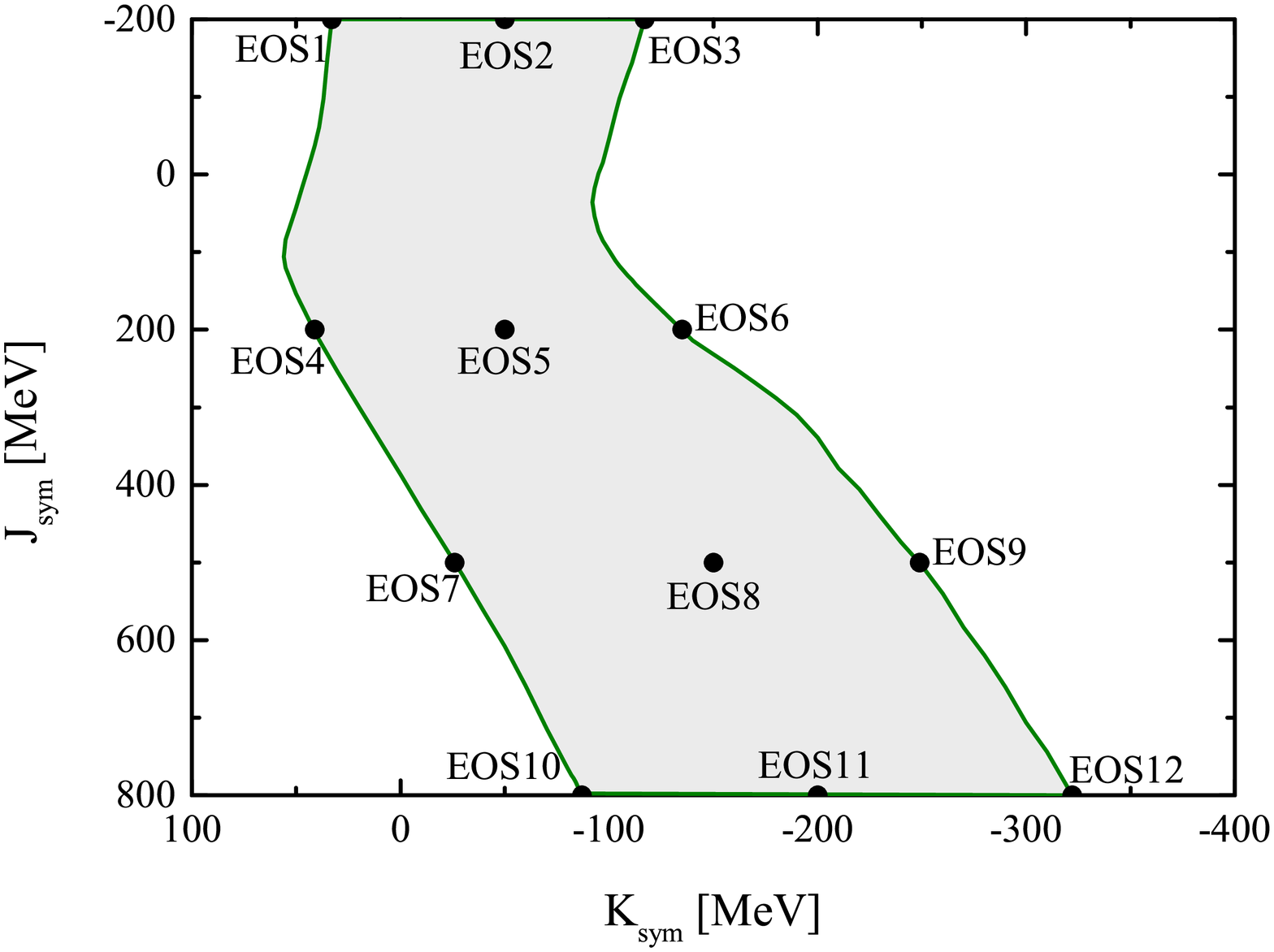}\\
  \caption{Projection of the constrained causality surface on the $K_{\rm sym}-J_{\rm sym}$ plane determined by the crosslines of the constant surfaces shown in Figure \ref{Constraints}. The shadowed range corresponds to the parameters allowed. The black dots indicate the EOS parameter sets chosen to calculate properties of pulsars on the causality surface.}\label{KsymJsym}
\end{figure}

Shown in Figure\ \ref{Constraints} are the tightest constraints on the 3D high-density EOS parameter space from inverting the indicated radii and tidal deformability of canonical NSs \citep{Zhang2020} as well as the causality condition and NSs' {\it minimum} maximum mass of M=2.14 M$_\odot$. The latter is the mass of PSR~J0740+6620 \citep{Mmax}. It is the confirmed most massive NS observed so far. So all acceptable EOSs have to be stiff enough to predict a mass-radius curve with a maximum at least as high as 2.14 M$_\odot$. Considering all possibly more massive NSs in the universe, 2.14 M$_\odot$ is the minimum maximum mass of acceptable EOSs. The surface labeled as M=2.14 M$_\odot$ in Figure\ \ref{Constraints} collects all EOSs that predict a NS maximum mass of 2.14 M$_{\odot}$. It limits the EOS space from below, while the upper bound is from the causality surface (blue) on which the speed of sound equals the speed of light ($v^2_s=dP/d\epsilon=c^2$) at the central density of the most massive NS supported by the nuclear pressure at each point with the specific EOS there \citep{Zhang19}.
Both surfaces are strongly controlled by the SNM EOS parameter $J_0$. As expected, these two surfaces are also significantly influenced by the high-density symmetry energy especially when the $E_{\rm{sym}}(\rho)$ becomes super-soft with negative values of $K_{\rm{sym}}$ and/or $J_{\rm{sym}}$.  For example, with the super-soft $E_{\rm{sym}}(\rho)$, to support the same NS maximum mass of M=2.14 M$_\odot$, the necessary value of $J_0$ has to become higher as one expects.
\begin{deluxetable*}{cccccccccc}
\tablenum{1}
\tablecaption{The labels and parameter sets of 12 EOSs on the bounded causality surface shown in Figure \ref{Constraints} and Figure \ref{KsymJsym}, the resulting maximum mass $M_{\rm TOV}$ of non-rotating NSs,  the maximum mass $M_{\rm RNS}$ of NSs rotating at the Kepler frequency $f_K$, the equatorial radius $R_{\rm RNS}$ of the NS with $M_{\rm RNS}$, the equatorial radius $R_{2.5}$ of the NS with mass 2.50 M$_\odot$ rotating at $f_{2.5}$,
the minimum frequency $f_{2.5}$ that can rotationally support an NS with mass 2.50 M$_\odot$, the ratio $f_{2.5}/f_K$, and the minimum spin parameter $\chi_{2.5}$ necessary to rotationally support the NS with mass 2.50 M$_\odot$.}\label{Table1}
\tablewidth{0pt}
\tablehead{
& \colhead{$(K_{\rm sym}, J_{\rm sym}, J_0)$} & \colhead{$M_{\rm TOV}$} & \colhead{$M_{\rm RNS}$} & \colhead{$R_{\rm RNS}$} &
\colhead{$R_{2.5}$} & \colhead{$f_{2.5}$} & \colhead{$f_{2.5}/f_K$}  & \colhead{$\chi_{2.5}$} \\
&\colhead{(MeV)} & \colhead{($M_\odot$)} & \colhead{($M_\odot$)} & \colhead{(km)} & \colhead{(km)} & \colhead{(Hz)} & &}
\startdata
EOS1  & (33, -200, 112.5)   & 2.39 & 2.87 & 14.77 & 11.92 & 971  & 0.578 & 0.375 \\
EOS2  & (-50, -200, 193.2)  & 2.29 & 2.73 & 14.47 & 12.83 & 1318 & 0.781 & 0.550 \\
EOS3  & (-117, -200, 225.2) & 2.14 & 2.53 & 13.56 & $-$   & $-$  & $-$   & $-$   \\
EOS4  & (41, 200, -69.6)    & 2.30 & 2.77 & 15.04 & 12.86 & 1217 & 0.757 & 0.516 \\
EOS5  & (-50, 200, -75.6)   & 2.28 & 2.73 & 14.47 & 12.80 & 1318 & 0.876 & 0.549 \\
EOS6  & (-135, 200, -199.4) & 2.14 & 2.55 & 13.67 & $-$   & $-$  & $-$   & $-$   \\
EOS7  & (-26, 500, -68.1)   & 2.33 & 2.80 & 15.30 & 12.67 & 1145 & 0.726 & 0.473 \\
EOS8  & (-150, 500, -88.0)  & 2.30 & 2.76 & 14.67 & 12.46 & 1265 & 0.759 & 0.514 \\
EOS9  & (-249, 500, -158.6) & 2.14 & 2.59 & 13.93 & $-$   & $-$  & $-$   & $-$   \\
EOS10 & (-87, 800, -65.2)   & 2.34 & 2.83 & 15.42 & 12.61 & 1073 & 0.683 & 0.444 \\
EOS11 & (-200, 800, -77.7)  & 2.34 & 2.83 & 14.95 & 12.36 & 1117 & 0.681 & 0.451 \\
EOS12 & (-322, 800, -184.1) & 2.14 & 2.62 & 14.49 & $-$   & $-$  & $-$   & $-$   \\
\enddata
\tablecomments{Though the maximum mass of neutron stars rotating at Kepler frequency for EOS3, EOS6, EOS9, and EOS11 is larger than 2.50 M$_\odot$, their maximum mass is too close to 2.50 M$_\odot$ for the RNS to output the $f_{2.5}$ sequences.}
\end{deluxetable*}
\begin{figure*}
\begin{center}
\resizebox{0.8\textwidth}{!}{
 \includegraphics{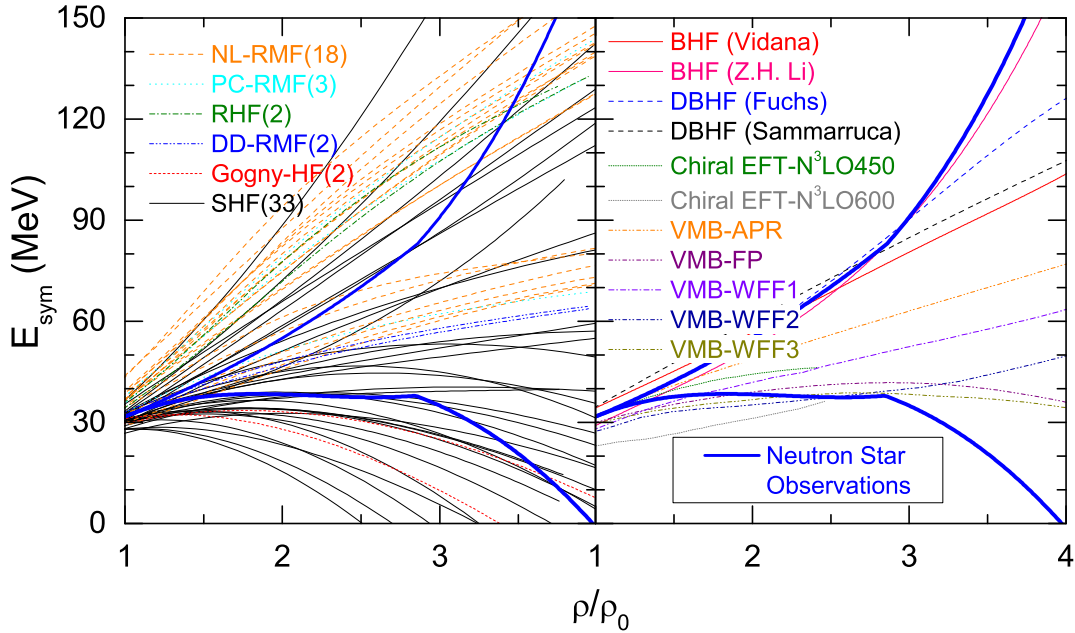}
}
\vspace{0.5cm}
\caption{Examples of the density dependence of nuclear symmetry energy predicted by nuclear many-body theories using different interactions, energy density functionals and/or techniques
in comparison with the constraining boundaries extracted from studying properties of neutron stars (the upper/lower thick-blue line corresponds to the left/right green boundaries shown in Figure \ref{KsymJsym}).}
\label{examples}
\end{center}
\end{figure*}

\begin{figure*}
  \centering
  \includegraphics[width=14cm]{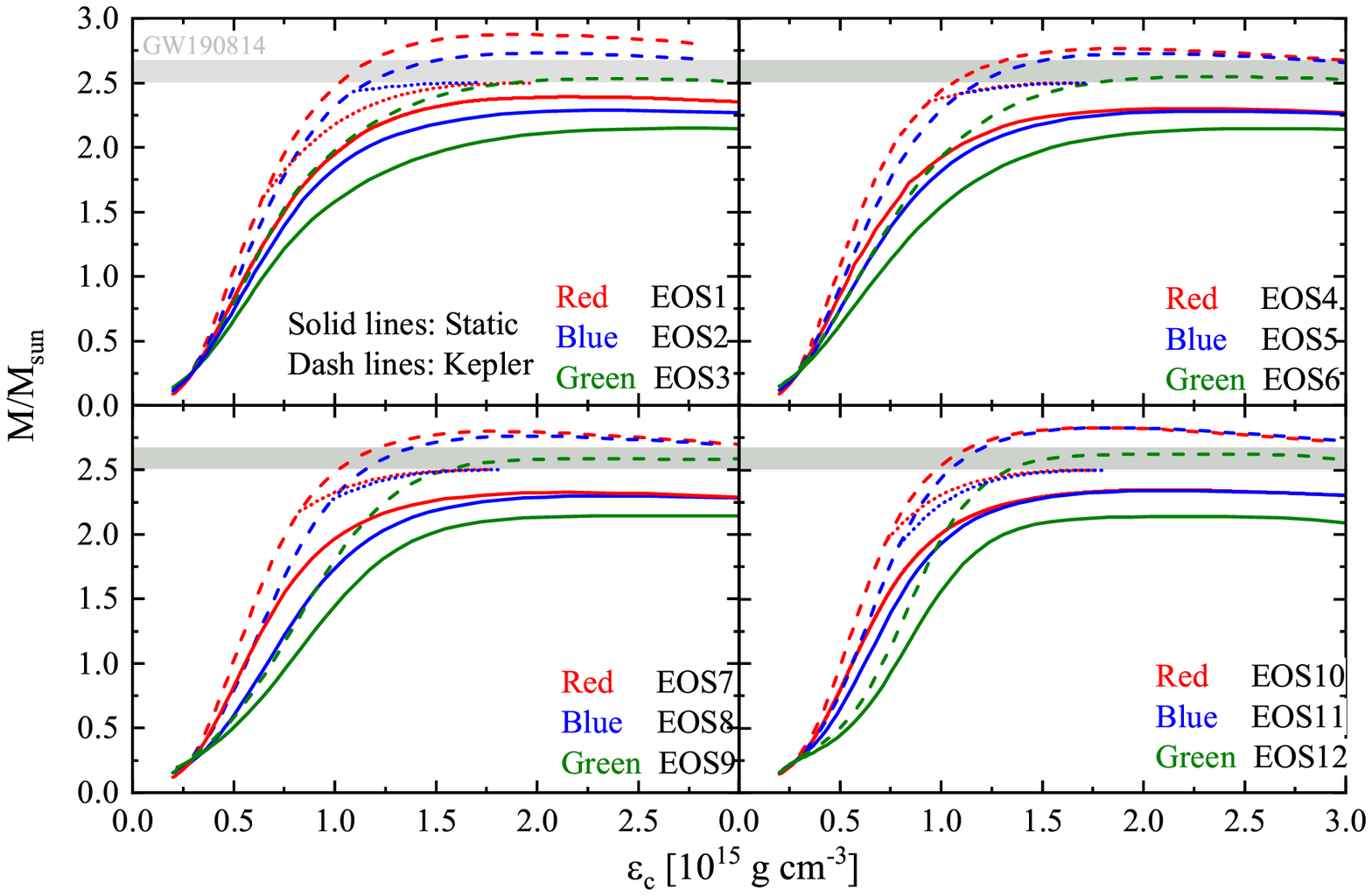}
    \includegraphics[width=14cm]{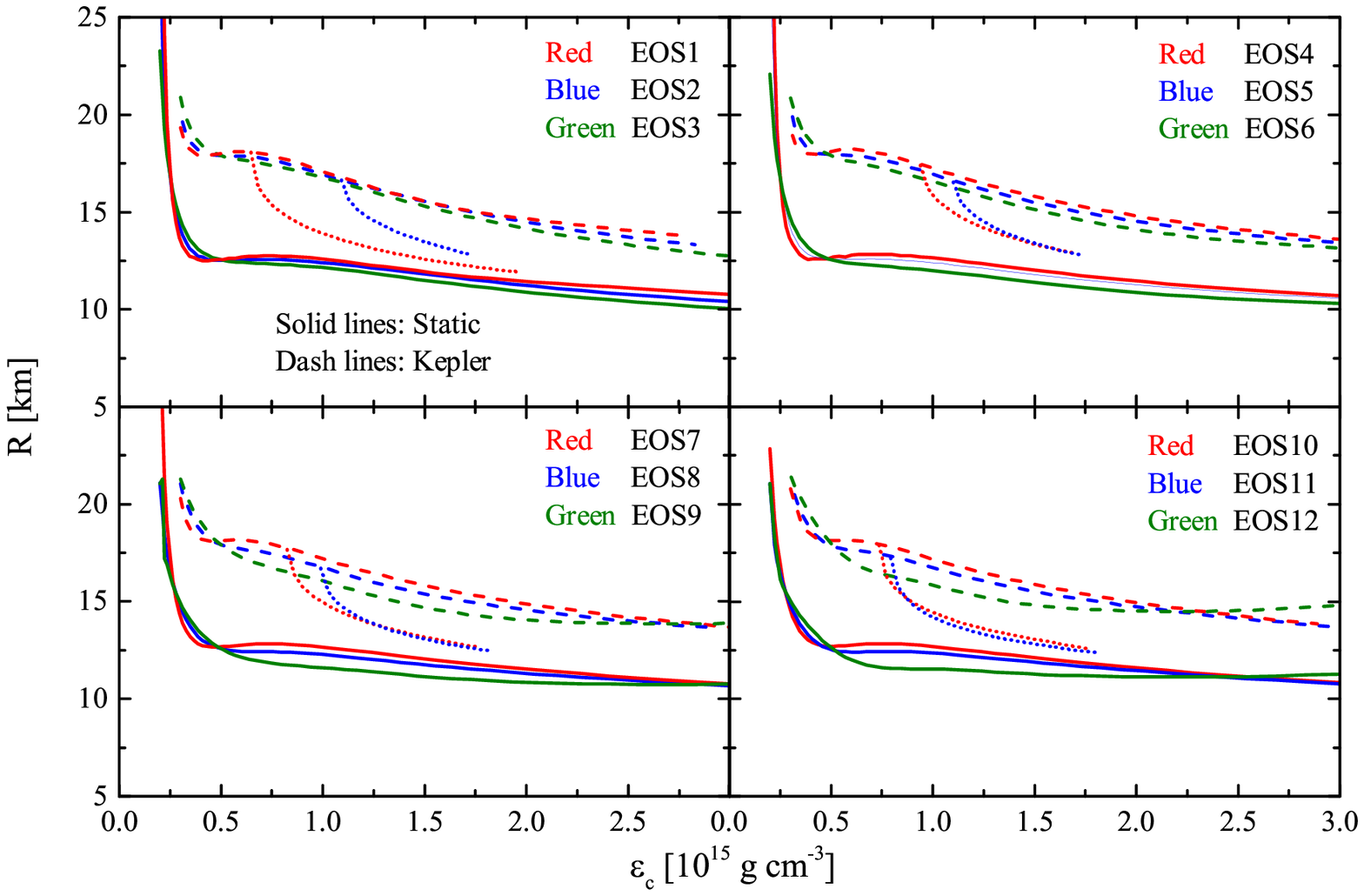}
  \caption{The mass (upper) and equatorial radius (lower) of static (solid lines) and rapidly rotating neutron stars as functions of their central energy density for the EOS parameter sets marked in Figure \ref{KsymJsym} and listed in Table \ref{Table1}. The neutron stars rotating at their respective Kepler frequencies and the minimum frequency $f_{2.5}$ that can rotationally support a neutron star with mass $2.50$ M$_\odot$ are shown as dashed lines and dotted lines, respectively. The reported mass $2.50-2.67$ M$_\odot$ of GW190814's secondary component is shown as gray bands.}\label{MRC}
\end{figure*}

We considered several reported radius and tidal deformability measurements, such as $10.62<R_{\rm{1.4}}< 12.83$ km from analyzing quiescent low-mass X-ray binaries \citep{Lattimer2014}, the dimensionless tidal deformability $70 \leq \Lambda_{1.4}\leq 580$ from the refined analysis of GW170817 data \citep{LIGO18}, the mass and radius of PSR J0030+0451 M$=1.44^{+0.15}_{-0.14}$ M$_\odot$ and R$=13.02^{+1.24}_{-1.06}$ km \citep{Miller2019} or M$=1.34^{+0.16}_{-0.15}$ M$_\odot$ and R$=12.71^{+1.19}_{-1.14}$ km \citep{Riley2019} from NICER. Both the upper and lower limits of radii from these measurements of canonical NSs are consistent. The ones shown in the Figure \ref{Constraints} provides the strongest constraint on the $K_{\rm sym}-J_{\rm sym}$ correlation. We notice that the lower radius boundary $R_{1.28} = 11.52$ km for M=1.28 M$_\odot$ from NICER is just outside the crossline between the causality surface and constant maximum mass surface of M=2.14 M$_\odot$.  It is known that the extraction of the lower limit of $\Lambda_{1.4}$ from GW170817 suffers from large uncertainties and is largely model dependent. The constant surface of $\Lambda_{1.4}$=70 is actually on the right of the $R_{1.28} = 11.52$ km surface, and the upper limit for the radius R$\leq 13.85$ km from NICER is on the left of the constant surface with $R_{\rm{1.4}}=12.83$ km, they are thus not shown here. The almost vertical surfaces of the radius and tidal deformability indicate that they are not much affected by the high density SNM EOS parameter $J_0$ but depend strongly on the high-density symmetry energy parameters $K_{\rm sym}$ and $J_{\rm sym}$.

The constraints of M = $2.14$ M$_\odot$ (green surface), $R_{1.4} = 12.83$ km (yellow surface), and causality condition (blue surface) together enclose the allowed high-density EOS parameter space in $K_{\rm sym}-J_{\rm sym}-J_0$. In particular, the causality surface determines the absolutely maximum mass $M_{\rm TOV}$ of non-rotating NSs. To find the minimum rotational frequency of GW190814's secondary if it is a pulsar, we focus on the constrained causality surface in the following discussions. Its left boundary is determined by its crossline with the $R_{1.4} = 12.83$ km (or the very close-by $\Lambda_{1.4} = 580$) surface,  while its right boundary is determined by its crossline with the M = $2.14$ M$_\odot$ surface. To be more clear, these crosslines are projected to the $K_{\rm sym}-J_{\rm sym}$ plane in Figure \ref{KsymJsym}. The shadowed range corresponds to the parameters allowed.

The astrophysical constraining boundaries on the high-density symmetry energy parameters shown in Figure \ref{KsymJsym} have significant impacts on both nuclear theories and experiments. As an illustration, we examine their impacts on theoretical predictions of nuclear symmetry energy at supra-saturation densities in Fig. \ref{examples}. In fact, essentially all two-body and/or three-body nuclear forces available in the literature have been used in one way or another in all available nuclear many-body theories to predict the density dependence of nuclear symmetry energy \citep{Tesym}. Shown in the left window are 60 representatives selected from 6 classes of totally over 520 energy density functional theories including the Relativistic Mean Field (RMF) using 3 different kinds (NL-RMF, PC-RMF and DD-RMF) of coupling schemes, Relativistic Hartree-Fock (RHF), Gogny Hartree-Fock (Gogny-HF) and Skyrme Hartree-Fock (SHF) with typical interactions. Detailed list of these interactions and models can be found in \citet{Chen2017}. Shown in the right window are 11 representatives of more microscopic and/or {\it ab initio} theories including the Brueckner Hartree Fock (BHF) \citep{LiZ08,Vid09}, Dirac-Brueckner Hartree Fock (DBHF) \citep{Fuc06,Kla06,Sam10}, Chiral Effective Field Theory (Chiral EFT) with 2 different high-momentum cut-offs \citep{Fra14} and the 3 versions (VMB-APR,VMB-FP and VMB-WFF) of the Variational Many Body (VMB) theories \citep{Fri81,WFF12,APR98} using different interactions. It is seen that in both windows the predictions diverge broadly. Interestingly, the astrophysical constraints from analyzing properties of NSs can already exclude many of the predictions especially those based on energy density functionals. While the constraints are still rather loose at densities above $2\rho_0$.  As we discussed earlier, this is mainly because the $J_{\rm{sym}}$ parameter controlling the symmetry energy at high densities above $2\rho_0$ are not constrained by properties of canonical NSs of masses around 1.4 M$_{\odot}$. Thus, the comparisons in Fig. \ref{examples} illustrate both the important impacts of NS observations on nuclear physics and the clear need to further constrain the high density behavior of nuclear symmetry energy, especially the $J_{\rm{sym}}$ parameter.

In the following, we choose 12 parameter sets (black dots) by varying the $J_{\rm sym}$ parameter from -200 to 800 MeV along and/or inside the boundaries. The specific values of the $K_{\rm sym}$, $J_{\rm sym}$, and $J_0$ parameters and the resulting properties of both non-rotating and uniformly rotating NSs are summarized in Table \ref{Table1}. It is particularly interesting and useful for the following discussions to note that the $M_{\rm TOV}$ on the bounded causality surface is between 2.14 and 2.39 M$_\odot$.

\begin{figure*}[htb]
\begin{center}
\resizebox{0.9\textwidth}{!}{
  \includegraphics{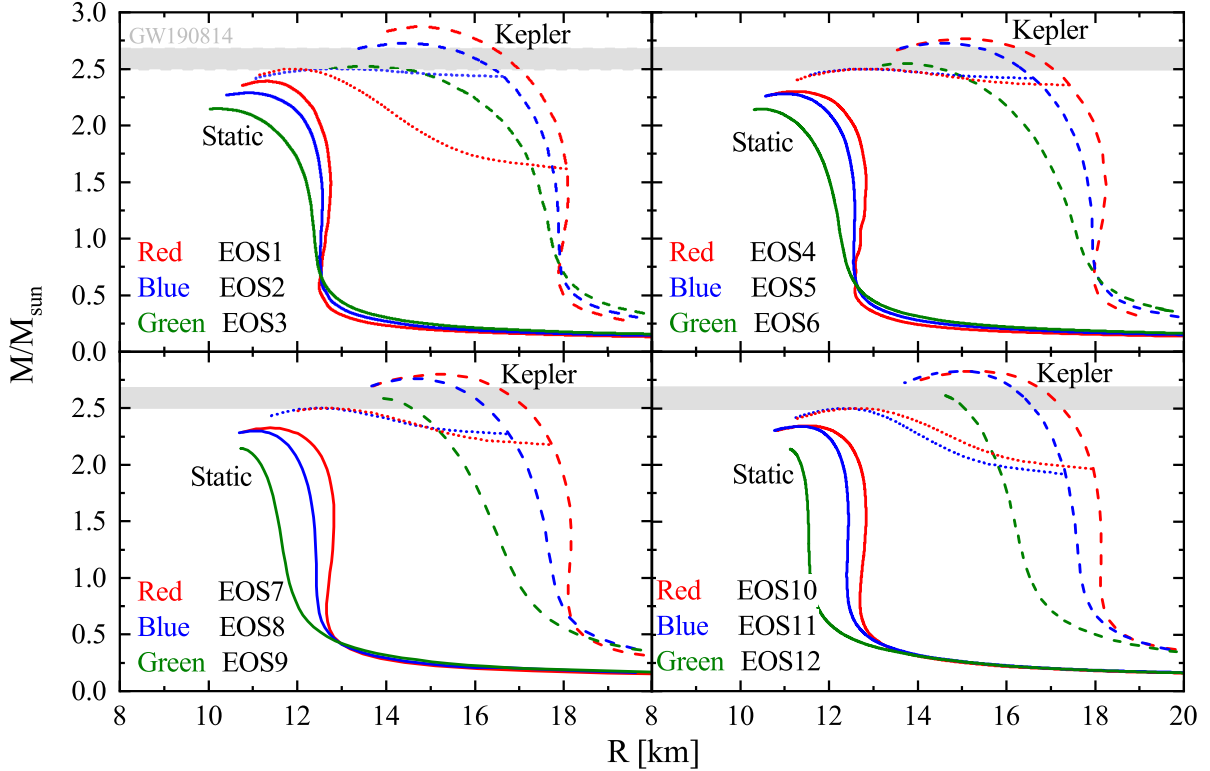}
   }
  \caption{Same as Figure \ref{MRC} but for the M-R relations of both static and rotating neutron stars.}\label{MR}
\end{center}
\end{figure*}

\section{Necessary observable properties of GW190814's secondary component as a super-fast pulsar}\label{prop}
Using Stergioulas's RNS code and the 12 EOSs on the causality surface discussed above, we now study the necessary properties for the GW190814's secondary component to be an NS. For technical and numerical details of the RNS code, we refer the readers to \citet{RNS} and its underlying physics to \citet{Komatsu1989,Cook1994,Stergioulas1995,Nozawa1998}.

For the purposes of this work, we examine the following NS rotational properties:
\begin{itemize}
  \item The mass-radius relations of fast pulsars with respect to those of non-rotating NSs, including the maximum mass $M_{\rm TOV}$ of non-rotating NSs, the pulsar maximum mass $M_{\rm RNS}$ at the Kepler frequency $f_K$ which is the maximum frequency that the gravitational attraction is still sufficient to keep matter bound to the pulsar surface
  \item The minimum frequency $f_{2.5}$ (and the ratio $f_{2.5}/f_K$) necessary to rotationally support a pulsar with mass 2.50 M$_\odot$ for a given EOS
 \item  The equatorial radius $R_{\rm RNS}$ of the pulsar with mass $M_{\rm RNS}$, the equatorial radius $R_{2.5}$ of the pulsar with mass 2.50 M$_\odot$ and frequency $f_{2.5}$
 \item The dimensionless spin parameter $\chi=J/M^2$ where $J$ is the angular momentum of the pulsar and its minimum value $\chi_{2.5}$ necessary to support the pulsar with mass 2.50 M$_\odot$.
\end{itemize}

Shown in Fig. \ref{MRC} are the mass (upper) and equatorial radius (lower) of static (solid lines) and rapidly rotating neutron stars as functions of their central energy density for the EOS parameter sets marked in Figure \ref{KsymJsym} and listed in Table \ref{Table1}. The neutron stars rotating at their respective Kepler frequencies and the minimum frequency $f_{2.5}$ that can rotationally support a neutron star with mass $2.50$ M$_\odot$ are shown as dashed lines and dotted lines, respectively. The reported mass $2.50-2.67$ $M_\odot$ of GW190814's secondary component is shown as gray bands.
The corresponding mass-radius relations are shown in Figure \ref{MR} and the resulting values of $M_{\rm TOV}$, $M_{\rm RNS}$, $R_{\rm RNS}$, $R_{2.5}$, $f_{2.5}$, the ratio $f_{2.5}/f_K$, and $\chi_{2.5}$
are summarized in Table \ref{Table1}.  Several interesting observations can be made from these results. We discuss the most important physics points in the following.

\begin{enumerate}

\item
The rotational effects on the mass and radius as well as their correlations are consistent with previous findings in the literature. Most interestingly, while the maximum M$_{\rm TOV}$ is 2.39 $M_\odot$ for the EOSs allowed by
the existing astrophysical observations and terrestrial experiments as we discussed in the previous section, rotations at frequencies much below the Kepler frequencies can readily bring the NS maximum mass to be above 2.50 M$_\odot$. This seemingly trivial result obtained from the well established theory/code for pulsars using the most conservative EOSs without introducing any new physics is important for the current debate whether the secondary component of GW190814 is an NS, a BH or a more exotic object. As we shall discuss in the following, all properties of GW190814's secondary component as a super-fast pulsar are consistent with expectations based on known physics. Thus, all together these lead firmly to our main conclusion that the secondary component of GW190814 is simply a super-fast pulsar rather than a BH or an exotic object.

\item While the M$_{\rm TOV}$ of the 12 EOSs are between 2.14 and 2.39 $M_\odot$, pulsars at their respective Kepler frequencies can easily sustain masses heavier than $2.50$ M$_\odot$. Of course, the maximum pulsar mass $M_{\rm RNS}$ depends sensitively on the EOS and the corresponding $M_{\rm TOV}$. With the stiffest EOS possible, i.e., the EOS1 with $M_{\rm TOV}=2.39$ M$_\odot$, the $M_{\rm RNS}=2.87$ M$_\odot$, while with the soft EOSs including EOS3, EOS6, EOS9, and EOS12 on the right boundary of the allowed EOS space shown in Figure \ref{KsymJsym} that is determined by the causality condition and the $M=2.14$ M$_\odot$ surface, the $M_{\rm RNS}$ are slightly larger than 2.50 M$_\odot$ but less than 2.67 M$_\odot$.  Consequently, for these soft EOSs all with the same $M_{\rm TOV}=2.14$ M$_\odot$, the minimum frequency $f_{2.5}$ necessary to rotationally support a pulsar with mass 2.50 M$_\odot$ should be only slightly smaller than their $f_K$ values. For this reason, the RNS code does not give the $f_{2.5}$ pulsar sequences with the EOS3, EOS6, EOS9, and EOS12.

\begin{figure}
  \centering
  \includegraphics[width=8cm]{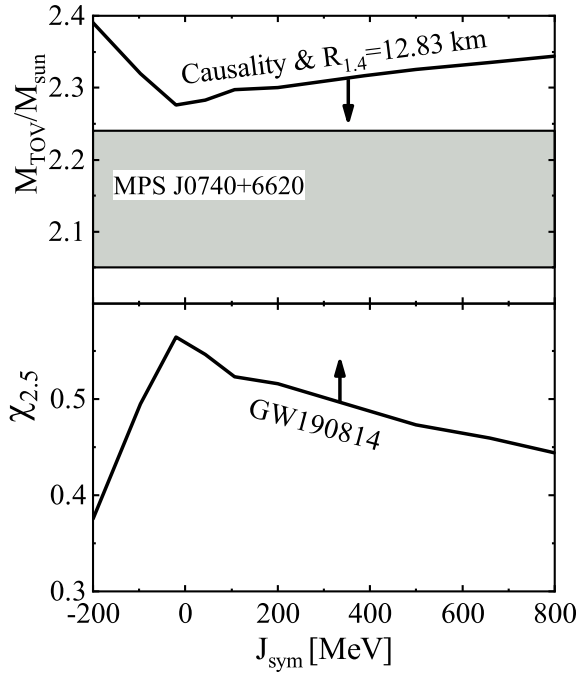}\\
  \caption{The maximum mass $M_{\rm TOV}$ of non-rotating NSs (upper window) and the minimum spin parameter $\chi_{2.5}$ of pulsars with the frequency $f_{2.5}$ (lower window) as functions of the high-density symmetry energy parameter $J_{\rm sym}$. The currently observed NS maximum mass $M=2.14^{+0.10}_{-0.09}$ M$_\odot$ (68\% confidence level) of MSR J0740+6620 is shown in the upper panel. The arrows indicate the conditions for GW190814's secondary to be a super-fast pulsar.}\label{Mchi25}
\end{figure}

\item The mass range on the mass-radius curve with a constant frequency becomes very narrow at higher frequencies \citep[see, e.g.,][for more detailed examples]{Haensel2008,Plamen08}. Indeed, the pulsar sequences at $f_{2.5}$ shown with the dashed and dotted lines are very flat. As expected, the stiffest EOS needs the lowest value of $f_{2.5}$. Thus,
as the stiffest EOS allowed, the EOS1 sets the lower limit of $f_{2.5}$ to $f_{2.5}>971$ Hz. Since the frequency of XTE J1739- 285 \citep{Kaaret2007} at 1122 Hz was not confirmed, the $f_{2.5}$ is higher than the confirmed highest frequency 716 Hz of PSR J1748-2446ad \citep{Hessels2006}. But it is still much less than the corresponding Kepler frequency $f_k$ with $f_{2.5}/f_K$=0.578. Obviously, the possibility for GW180814's secondary as a super-fast pulsar or even the fastest one ever found \citep{Most} cannot be excluded. The critical task is then to get more observational information about the secondary's spin.

\item  The minimum value of $\chi_{2.5}$ corresponding to the minimum $f_{2.5}$ is $0.375$ with the stiffest EOS, namely the EOS1. Since the fixed frequency pulsar sequences cannot be calculated with the RNS code when the $f_{2.5}$ approaches the Kepler frequency as we discussed above,  the upper boundary of $\chi_{2.5}$ is not determined here. However, it should be smaller than the maximum spin parameter $\chi_{\rm max}$, which is around $0.6-0.7$ and model-independent \citep{Friedman1992,Lo2011}. As shown in Figure\ \ref{Constraints}, the causality surface goes downwards towards its crossline with the 2.14 M$_\odot$ surface, namely, the EOS becomes softer with the decreasing $K_{\rm sym}$ when the $J_{\rm sym}$ is fixed. As a result, as shown in Table \ref{Table1}, the $M_{\rm TOV}$, $f_{2.5}$, and $\chi_{2.5}$ all decrease correspondingly. Thus, the left boundary of the projected EOS space shown in Figure \ref{KsymJsym} provides the lower boundary of $\chi_{2.5}$ and the upper boundary of $M_{\rm TOV}$. As shown in Figure\ \ref{Constraints}, this is the boundary set by the crossline between the causality surface and the surface with a constant radius of $R_{\rm{1.4}}< 12.83$ km.

\item The maximum mass $M_{\rm TOV}$ of non-rotating NSs (upper window) and the corresponding minimum spin parameter $\chi_{2.5}$ of pulsars with the frequency $f_{2.5}$ (lower window) are shown in Figure \ref{Mchi25} as functions of the parameter $J_{\rm sym}$. As we discussed earlier, the latter controls the behavior of nuclear symmetry energy at densities above $2\rho_0$. It is currently considered as the most uncertain parameter of the EOS of super-dense neutron-rich nucleonic matter \citep{Li17}. For a comparison, the mass $M=2.14^{+0.10}_{-0.09}$ M$_\odot$ (68\% confidence level) of MSR J0740+6620 is also shown in the upper panel. The arrows indicate the conditions for GW190814's secondary component to be a super-fast pulsar with its minimum spin parameter $\chi_{2.5}$. Combining the information from this plot and the constrained EOS parameter space shown in Figure \ref{Constraints}, clearly all the EOSs in the whole space between the causality surface and the $M=2.14$ M$_\odot$ surface can support pulsars as heavy as  2.50 M$_\odot$ if they rotate with varying minimum frequencies higher than 971 Hz depending on the symmetry energy of super-dense neutron-rich nuclear matter. This further illustrates the importance of better constraining the latter with terrestrial experiments and/or astrophysical observations.

\item The stiffest EOS, EOS1 ($K_{\rm sym}=33$ MeV, $J_{\rm sym}=-200$ MeV and $J_0=112.5$ MeV) requires the least spin parameter $\chi_{2.5}=0.375$. The corresponding $M_{\rm TOV}$=2.39 M$_\odot$ is a little higher than the $M_{\rm TOV}$=2.3 M$_{\odot}$ adopted by \citet{Most} from analyzing GW170817. Using the latter and assuming the radius of GW190814's secondary is $13$ km, they extracted a range of $0.49<\chi<0.68$ and $f>1140$ Hz for the spin parameter based on the universal relations of masses and spin parameters \citep{Breu}. Our results are qualitatively consistent and the quantitative difference can be well understood from the differences in the $M_{\rm TOV}$ and the pulsar radius used. In fact, as shown in Table \ref{Table1}, if we restrict the EOSs to the ones giving approximately $M_{\rm TOV}$=2.30 M$_\odot$ and $R_{2.5}=13$ km, our numerical results are even in better agreement.

\item There are some longstanding and interesting issues regarding the stability of fast pulsars \citep[see, e.g.,][]{Hessels2006,Haensel2008}, such as the r-mode instability in the cores of fast pulsars \citep[see, e.g.,][]{Lin98,Owen98,And2,Levin} that may happen at frequencies much lower than the Kepler frequency. The r-mode instability window depends strongly on the core temperature and its transport properties as well as the coupling with and structure of the crust. Its calculation is still very model dependent and relies on many poorly known properties of NS matter. For instance, it has been shown by \citet{DHW} and \citet{ISAAC12} that both the Kepler frequency $f_K$ and the boundaries of the r-mode instability window in the frequency-temperature plane have significant dependencies on nuclear symmetry energy. The separation between the $f_K$ and the critical frequency $f_r$ above which the r-mode instability occurs is strongly temperature dependent. How the minimum frequency $f_{2.5}$ for the GW190814's secondary component to be a super-fast pulsar compares with the critical r-mode instability frequency $f_r$ is an interesting question for future studies.

\end{enumerate}

\begin{figure}
  \centering
  \includegraphics[width=8cm]{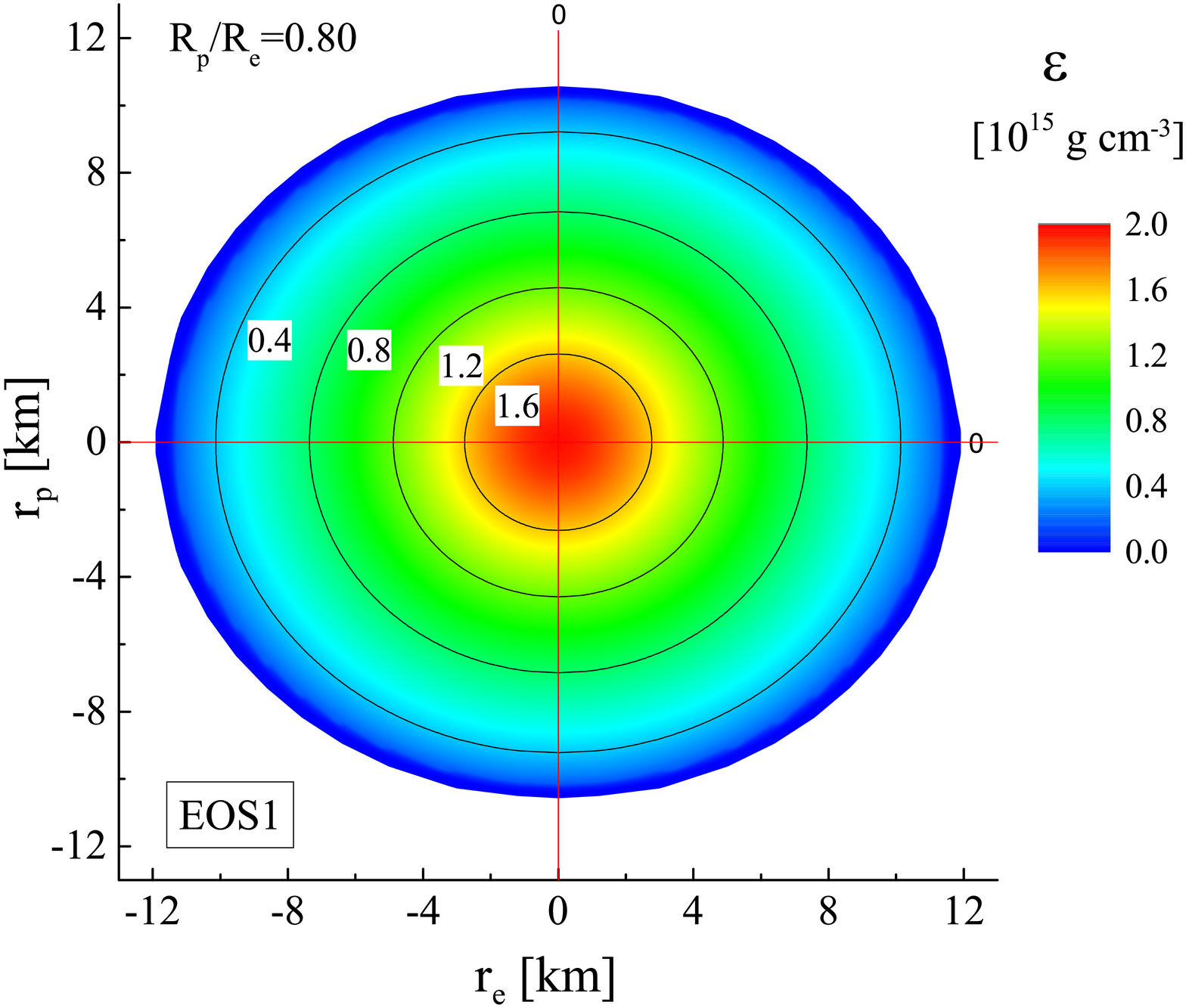}
    \includegraphics[width=8cm]{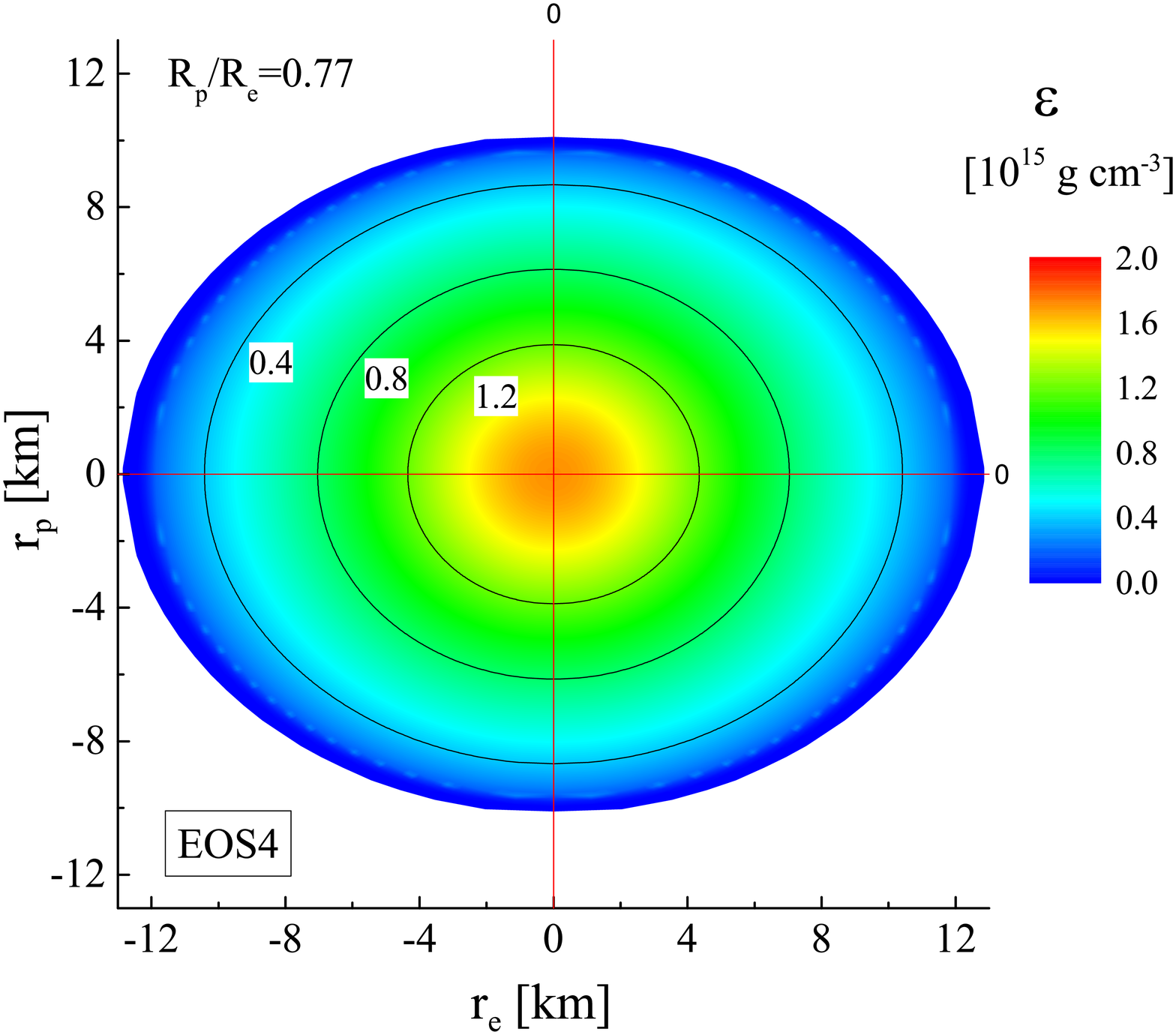}
  \caption{The energy density contours on the equatorial ($r_e$)-polar ($r_p$) plane with the EOS1 (upper) and EOS4 (lower) at the minimum frequency $f_{2.5}$ to rotationally support NSs with a mass of $M=2.50$ M$_\odot$, respectively.}\label{profile1}
\end{figure}

\begin{figure}
  \centering
  \includegraphics[width=9cm]{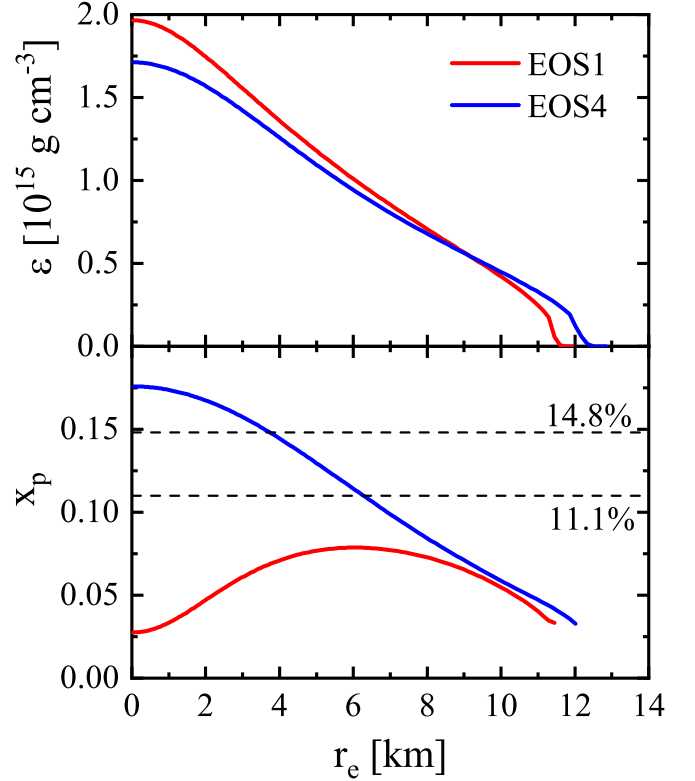}
    \caption{The radial profiles of energy density (upper) and proton fraction (lower) along the equatorial radius with EOS1 and EOS4, respectively. The horizontal dashed lines correspond to the threshold proton fraction above which the direct URCA cooling of protoneutron stars can happen in the $npe\mu$ matter, see the text for details.}\label{profile2}
\end{figure}

\section{Internal properties of GW190814's secondary component as a super-fast pulsar}
Besides the observational properties discussed above, it is also interesting to examine the corresponding internal properties of GW190814's secondary component as learning about them is the ultimate goal of all NS observations.
We have studied the profiles of the energy density $\epsilon$ and the proton fraction $x_p$ for all 12 EOSs considered. For comparisons, we present and discuss results with the EOS1 and EOS4. As listed in Table \ref{Table1}, the EOS1 has the highest $M_{\rm TOV}$ of 2.39 M$_\odot$ thus the lowest $f_{2.5}=971$ Hz necessary to rotationally support NSs with a mass of $M=2.50$ M$_\odot$,
while the EOS4 has almost the lowest $M_{\rm TOV}$ of 2.30 M$_\odot$ but the highest $f_{2.5}=1217$ Hz along the mass and spin boundaries shown in Fig. \ref{Mchi25}. Moreover, because the EOS1 has $J_{\rm sym}=-200$ MeV but EOS4 has $J_{\rm sym}=+200$ MeV while they have approximately the same $K_{\rm sym}$, they represent respectively the super-soft and stiff symmetry energy functionals at densities above $2\rho_0$ as shown in Fig. \ref{Esym}.  As discussed in Section \ref{EOSmodel}, one distinguished feature of our NS EOS-generator is the explicit isospin dependence and the ability to keep tracking the composition of NSs. Here we shall examine the proton fraction and its potential impact on fast cooling through the direct URCA process \citep{LPPH} with the EOS1 and EOS4.

Shown in Fig.\ \ref{profile1} are the energy density contours on the equatorial ($r_e$)-polar ($r_p$) plane with the EOS1 (upper) and EOS4 (lower) at the minimum frequency $f_{2.5}$ to rotationally support NSs with a mass of $M=2.50$ M$_\odot$, respectively. The EOS4 predicts an equatorial radius of about 0.9 km larger than that with the EOS1 as one expects since the EOS4 has a higher frequency $f_{2.5}=1217$ Hz. However, the ratio $R_p/R_e$ of polar over equatorial radius is only slightly different by about 4\% with the two EOSs. The EOS1 having the highest $M_{\rm TOV}$ of 2.39 M$_\odot$ also has the highest central energy density of about $2\times 10^{15}$ g/cm$^3$. The energy density decreases gradually towards the surface. This feature is shown more quantitatively in the upper window of Fig.\ \ref{profile2}. It is seen that the difference in energy density with the two EOSs occurs mostly in the central areas of NSs. 

Shown in the lower window of Fig.\ \ref{profile2} are the profiles of the proton fraction $x_p$. It is clearly seen that the NS with EOS1 is much more neutron-rich (proton-poor) than the one with the EOS4. In fact, the core of the NS with EOS1 is almost made of purely neutrons ($x_p\approx 0.025$). This is what one expects based on the discussions about the density dependence of symmetry energy in Fig.\ \ref{Esym}. Again, due to the $E_{\rm{sym}}(\rho)\cdot \delta^2$ term in the average nucleon energy in neutron-rich matter of Eq. (\ref{eos}), a super-soft (low value) symmetry energy with $J_{\rm sym}=-200$ MeV in the EOS1 makes the corresponding $\delta$ at $\beta$ equilibrium close to its maximum value of 1 at densities  above about $3.5\rho_0$. 

It is well known that the proton fraction is the most critical quantity determining the cooling mechanisms of protoneutron stars and the related neutrino emissions \citep{LPPH}. In the $npe\mu$ matter, the threshold proton fraction $x^{DU}_p$ enabling the fast cooling through the direct URCA process (DU) is
\begin{equation}\label{xdu}
x^{DU}_p=1/[1+(1+x_e^{1/3})^3]
\end{equation}
with $x_e\equiv \rho_e/\rho_p$ between 1 and 0.5 leading to a $x^{DU}_p$ between 11.1\% to 14.8\%  \citep{Kla06}. As indicated in the lower window of Fig.\ \ref{profile2}, the EOS4 allows the direct URCA in a large region of the core but the EOS1 completely forbids it. This has significant implications for some NS observables, such as the neutrino flux and surface temperature. In turn, observational data of these observables will allow us to probe the high-density behavior of nuclear symmetry energy. Hopefully, future analyses of GW190814 or similar events will make this possible.

\section{Summary and Conclusion}
Using Stergioulas's RNS code for investigating fast pulsars with EOSs on the causality surface and allowed by all known constraints from both nuclear physics and astrophysics, we found that the GW190814's secondary component can be a super-fast pulsar as long as it rotates faster than 971 Hz about 42\% below its Kepler frequency. There is a large high-density EOS parameter space below the causality surface permitting pulsars heavier than 2.50 M$_{\odot}$ if they can rotate even faster with varying critical frequencies depending strongly on the high-density behavior of nuclear symmetry energy. 

Interestingly, it was suggested very recently that the secondary was born as a NS where a significant amount of the
supernova ejecta mass from its formation remained bound to the binary due to the presence of the massive BH companion \citep{Harvard}. In this model, very high spin angular momentum, such as the one we found here necessary to rotationally support GW190814's secondary as a super-fast pulsar, could be supplied through the circumbinary accretion disk \citep{Harvard,Harvard2}.
To rule out completely the possibility for the GW190814's secondary component as a super-fast pulsar, it is critical to observationally constrain its spin properties. To better understand the properties of super-fast pulsars it is important to further constrain the high-density behavior of nuclear symmetry energy with both astrophysical observations and/or terrestrial nuclear experiments. In turn, if confirmed as the most massive and fastest pulsar observed so far, the cooling curve and/or the associated neutrino emission of GW190814's secondary will provide a great opportunity to further probe the symmetry energy of super-dense neutron-rich nuclear matter.

\section{Acknowledgement}
We would like to thank Dr. Ronaldo Vieira Lobato for useful communications. This work is supported in part by the U.S. Department of Energy, Office of Science, under Award Number DE-SC0013702, the CUSTIPEN (China-U.S. Theory Institute for Physics with Exotic Nuclei) under the US Department of Energy Grant No. DE-SC0009971, the China Postdoctoral Science Foundation under Grant No. 2019M652358, and the Fundamental Research Funds of Shandong University under Grant No. 2019ZRJC001.


\end{document}